\documentclass[fleqn,usenatbib]{mnras}

\usepackage[T1]{fontenc}
\usepackage{ae,aecompl}
\usepackage{subfigure}
\usepackage{xcolor}
\usepackage[normalem]{ulem}
\usepackage{microtype}

\newcommand{\pks}{{PKS 0558$-$504}}

\newcommand{\tm}{\emph{$\tau_{\rm jav}$\,}}

\newcommand{\w}{W2}
\newcommand{\ww}{W1}
\newcommand{\m}{M2}
\newcommand{\py}{PyceCREAM}

\newcommand{\orcid}[1]{\textsuperscript{\href{http://orcid.org/#1}{
\hskip2pt\includegraphics[width=8pt]{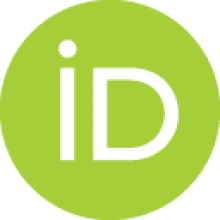}}}}
\usepackage{graphicx}	
\usepackage{amsmath}	
\usepackage{amssymb}	

\title[PKS 0558$-$504]{Revisiting the Continuum Reverberation Lags in the AGN PKS 0558$-$504}
\author[Gonzalez-Buitrago et al.]{D. H. Gonz\'{a}lez-Buitrago\orcid{0000-0002-9280-1184},$^{1}$\thanks{E-mail: dgonzalez@astro.unam.mx}
J.~V. Hern\'andez Santisteban\orcid{0000-0002-6733-5556},$^{2}$
A. J. Barth\orcid{0000-0002-3026-0562},$^{3}$
\newauthor
E. Jimenez-Bail\'on,$^{1}$
Yan-Rong Li\orcid{0000-0001-5841-9179},$^{5}$ 
Ma. T. Garc\'ia-D\'iaz\orcid{0000-0002-9772-5555},$^{1}$
A. Lopez Vargas$^6$,
\newauthor
and
M. Herrera-Endoqui$^{1}$
\\
$^{1}$Universidad Nacional Aut\'onoma de M\'exico, Instituto de Astronom\'ia, AP 106,  Ensenada 22860, BC, M\'exico\\
$^{2}$SUPA School of Physics \& Astronomy, University of St Andrews, North Haugh, St Andrews KY16 9SS, Scotland, UK\\
$^{3}$Department of Physics and Astronomy, 4129 Frederick Reines 
Hall, University of California, Irvine, CA, 92697-4575, USA\\
$^4$ Instituto de Astronom{\'\i}a, Universidad Nacional Aut\'onoma de M\'exico, Apartado Postal 70-264, 04510 CDMX, Mexico\\
$^5$ Key Laboratory for Particle Astrophysics, Institute of High Energy Physics, Chinese Academy of Sciences, 19B Yuquan Road,\\ Beijing 100049, China.\\
$^6$Facultad de Ciencias Exactas, Universidad Aut\'onoma del Estado de Baja California,
km 6.7 Carr. Transpeninsular 3917,\\22860 Ensenada, B.C., Mexico
}

\date{Accepted 2022 July 3. Received  2022 June 5; in original form 2022 February 21}
\pubyear{2021}

\begin{document}
\label{firstpage}
\pagerange{\pageref{firstpage}--\pageref{lastpage}}
\maketitle

\begin{abstract}
We present a revised analysis of the photometric reverberation mapping campaign of the narrow-line Seyfert 1 galaxy \pks\, carried out with the \textit{Swift} Observatory during 2008--2010. Previously, Gliozzi et al.\ found using the Discrete Correlation Function (DCF) method that the short-wavelength continuum variations lagged behind variations at longer wavelengths, the opposite of the trend expected for thermal reprocessing of X-rays by the accretion disc, and they interpreted their results as evidence against the reprocessing model. We carried out new DCF measurements that demonstrate that the inverted lag-wavelength relationship found by Gliozzi et al.\ resulted from their having interchanged the order of the driving and responding light curves when measuring the lags. To determine the inter-band lags and uncertainties more accurately, we carried out new measurements with four independent methods. These give consistent results showing time delays increasing as a function of wavelength, as expected for the disc reprocessing scenario. The slope of the re-analysed delay spectrum appears to be roughly compatible with the predicted $\tau \propto \lambda^{4/3}$ relationship for reprocessing by an optically thick and geometrically thin accretion disc, although the data points exhibit a large scatter about the fitted power-law trend.

\end{abstract}



\section{Introduction}

Active Galactic Nuclei (AGN) are the most luminous persistent sources in the Universe ($\gtrsim10^{42}$ erg s$^{-1}$) \citep{Lyden69}. They represent the active stage of supermassive black hole growth \citep{Kormendy13} at the core of most galaxies as material reaches its vicinity often through an accretion disc \citep{SS76}.  
Detailed characterisation of these regions is challenging due to their small angular size proving inaccessible for direct imaging  \citep[except for the notable example of M87$^{*}$,][]{EHT19}. AGN are interestingly variable sources over a wide energy range on different timescales \citep{Matthews63, Smith63}. This variability and the associated timescales carry information of the emitting region sizes  as well as provide a window to measure fundamental properties of the supermassive black hole at its centre \citep{Shen15a,Shen16b,Yue18}, in a method known as reverberation mapping \citep[RM;][]{Blandford82, Peterson14}.

Continuum variability spanning X-ray through optical wavelengths provides important clues to the size and structure of the accretion disc in AGN. In the standard lamp-post reprocessing model \citep[e.g.,][]{Frank02}, photons originating in the X-ray emitting corona located above the central region of the accretion disc travel to the external parts of the accretion disc and are locally reprocessed into photons of longer wavelength, with a characteristic time delay, $\tau$, that depends on the light-travel time from the corona to the disc. For the temperature profile of a geometrically thin disc, this model predicts that the continuum lags from UV through optical wavelengths should follow $\tau\propto\lambda^{4/3}$ \citet{SS73}. Thus, by measuring the delay time of the different continuum bands (which probe different regions of the disc) it is possible to test this model and to map the size and temperature profile of the disc.
Previous continuum RM studies  \citep[e.g.,][]{Cackett2007,Edelson15,Edelson17,Edelson19,Cackett2020,hernandez2020} have shown that wavelength dependent measurements are broadly consistent with the predictions of the \citet{SS73} model where the average delay as a function of wavelength scales as $\tau\propto \lambda^{4/3}$. However, these studies have also found that disc size estimation are $\sim 3-4 $ times larger than expected \citep{Edelson19}, in agreement with microlensing observations \citep[e.g.,][]{Morgan2010}.

\pks\ (\emph{z}=0.1372) is a variable quasar on different time scales and over a wide energy range from X-rays to the near infrared and with frequent flares \citep{Gli07,Gli10}. Different studies have estimated its black hole mass through multiple methods obtaining results between $\sim (2-4) \times 10^8 M_{\odot}$ and with a super-Eddington luminosity, $L/L_{\rm Edd} = 1.7$ \citep{Gli10}.
In addition, radio observations revealed the existence of an extended and aligned structure characteristic of bipolar jets \citep{Gli10}, with properties analogous to Galactic stellar black holes. 
The first continuum RM study of \pks\ was carried out by \citet{Gli13} using data from the \textit{Neil Gehrels Swift Observatory} \citep[hereafter \textit{Swift},][]{Gehrels04} employing simultaneous UVOT \citep{Roming05} and XRT \citep{Burrows05} observations. While the X-ray, UV, and optical bands light curves show a strong correlation between them, \citet{Gli13} found that variations in the optical bands led (rather than lagged behind) the corresponding UV variations, and the UV led the X-ray variations. This result suggested that the optical bands are responsible for the variations observed in X-rays, very different from that predicted by the reprocessing model and in contrast to most continuum RM studies. \citet{Gli13} interpreted this result as due to  fluctuations in the disc that drive the variability and propagate from the outer disc (optical) to the internal regions of the disc (UV) and  corona (X-rays) \citep{Lyubarskii97,Arevalo08}. Given this puzzling behaviour, which exhibited the opposite lag-wavelength trend seen in other AGN, we revisited the measurement of the \pks\ reverberation lags with the goal of better understanding the origin of the inverted lag-wavelength relationship found by \citet{Gli13}.

In this work, we present a new analysis and result of the multi-band continuum RM of \pks. The paper is divided as follows: in Section~\ref{sec:observations} we describe the observations and data reduction procedures. We present a revised time series analysis of the \textit{Swift} data and new measurements of lag time in Sections~\ref{sec:TSA} and \ref{sec:cream}. We present a discussion of the revised analysis in Section~\ref{sec:discussion} and conclusions in Section~\ref{sec:conclusion}.

\section{Observations and data}
\label{sec:observations}

A long-duration multi-wavelength monitoring of \pks\ was carried out with {\em Swift} between September 9, 2008, and March 30, 2010, covering a total of 90 visits during this period, with approximately $\sim 2$ ks of exposure time per visit every week. Simultaneous observations were made at each visit with UVOT's 6 filters (\w, \m, \ww, U, B, and V), and with XRT, in which the window timing mode was used. We obtained the light curves of the X-ray, UV, and optical bands from the online data tables of \citet{Gli13}, where further details on the data collection and reduction process are described. These data include a correction for line-of-sight extinction of $E(B - V) = 0.044$ mag \citep{Schlegel98}.  The UVOT light curve data provided by \citet{Gli13} are given in magnitudes, and we converted the data to flux densities ($f_\lambda$) to carry out lag measurements.

\section{Time-series analysis}
\label{sec:TSA}

We revisited the time-series analysis of \pks\ to determine the time delays between the X-ray, UV, and optical bands, using multiple methods as consistency checks. In all cases, lags are measured relative to the \w\,$(\lambda1928$\,\AA) band as the reference or driving band.

\label{sec:measurements}

\subsection{Discrete Correlation Function}
\begin{figure*}
 	\includegraphics[trim=0.0cm 0.0cm 0.0cm 0.0cm, clip, width=15cm]{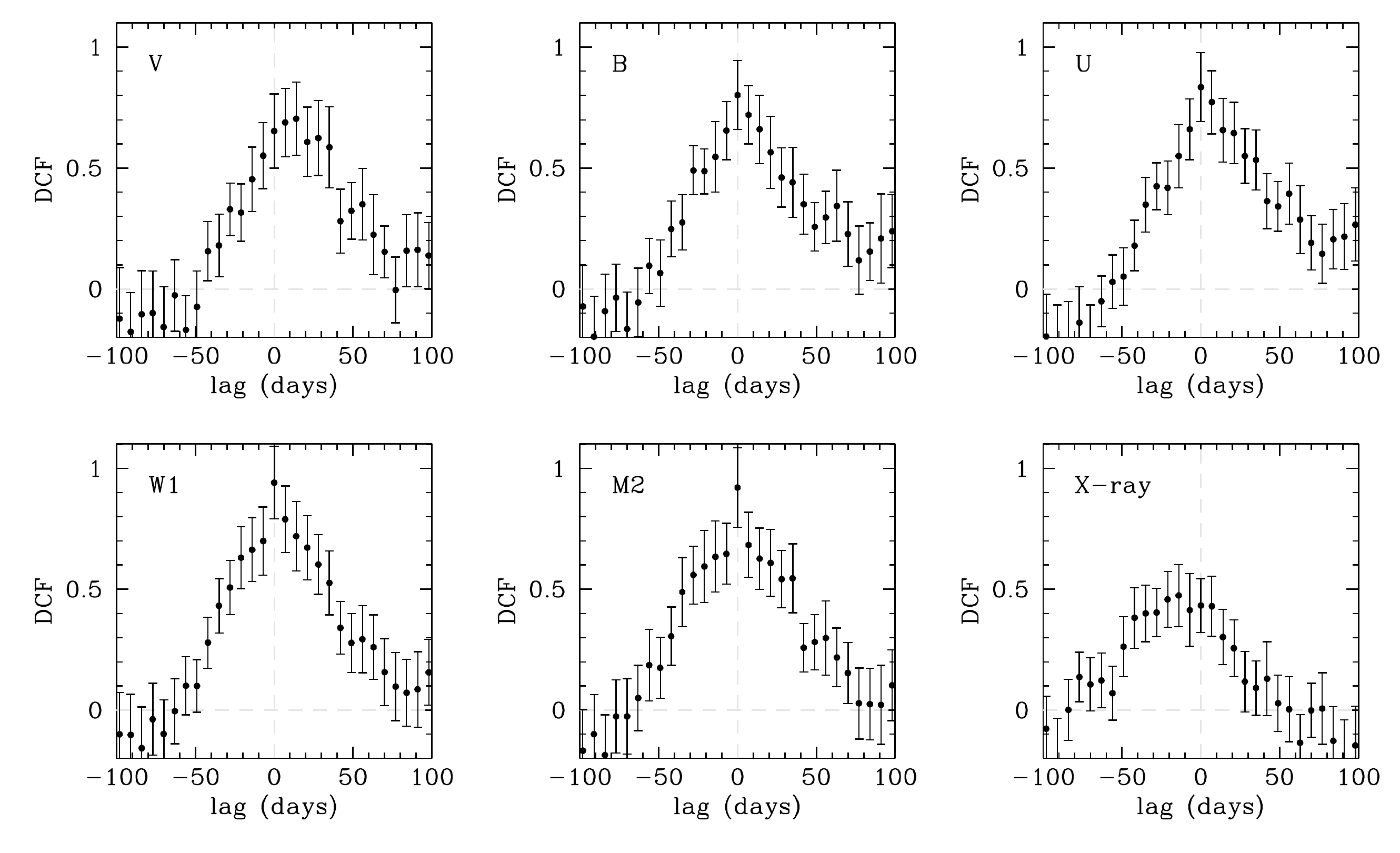}
\caption{Cross-correlation functions measured with the DCF method for the V, B, U, W1, M2, and X-ray bands. In each case, the \w\, band was used as the driving band.  The asymmetric structure in these CCFs, particularly the V and X-ray bands, shows the opposite shape seen in the CCFs in Figure 5 of \citet{Gli13}, indicating that their measurements must have inadvertently been carried out using the \w\, band as the responding rather than the driving band. The shape of these CCFs clearly illustrates that the V-band variations lag behind the \w, and \w\, lags behind the X-ray band.}
    \label{fig:dcf}
\end{figure*}

The lag measurements presented by \cite{Gli13} were done using the Discrete Correlation Function (DCF) method of \cite{Edelson88}. This method determines the cross-correlation function (CCF) between two unevenly sampled light curves by binning the CCF into regularly spaced temporal bins. With the goal of replicating the \cite{Gli13} lag measurements, we applied the DCF method to the \pks\ light curve data, measuring the CCF of each band relative to the \w\, band as the driving light curve. We followed the earlier measurement by using a 7-day bin size for the DCF. The resulting CCFs are shown in Figure \ref{fig:dcf}, and can be directly compared with the CCFs in Figure 5 of \citet{Gli13}. 

There are small differences in the point-to-point scatter between our CCFs and those of \citet{Gli13}. These may be attributed to using different implementations of the DCF algorithm although we are unable to point to a specific cause. Aside from these minor details, the shapes of our CCFs are largely the same as those presented by \citet{Gli13}, with one major overall difference: comparing our DCFs with those presented in Figure 5 of \citet{Gli13}, it is immediately clear that their CCFs are time-reversed versions of ours-- that is, mirror images reflected about $\tau=0$. This is most obvious in the V band, where the \citet{Gli13} CCF shows a clear negative lag while ours shows a positive lag, and in the X-ray band, where the \citet{Gli13} CCF indicates a positive lag while ours is negative. In the U and B bands, the CCFs are peaked near $\tau=0$ but slightly asymmetric, and our CCFs show the opposite sense of asymmetry from those of \citet{Gli13}. The CCFs for the W1 and M2 bands are peaked at zero lag and nearly symmetric.

Since our CCFs were measured using W2 as the driving continuum band, we conclude that the inverted lag-wavelength relationship found by \citet{Gli13} must have been the result of their having interchanged the order of the driving and responding bands when applying the DCF method. That is, they must have inadvertently used the W2 band as the responding band rather than the driving band when calculating the DCF, which would have the effect of time-reversing the CCF structure. Our updated DCF measurements demonstrate that the continuum lags in \pks\ actually behave in the usual, expected manner, with shorter-wavelength variations occurring first and longer-wavelength variations responding at later times.

\citet{Gli13} performed a Monte Carlo bootstrapping analysis, generating multiple randomised versions of the light curves and re-running the DCF to produce an ensemble of results to evaluate the lag of the CCF peak and its uncertainty. However, the DCF method can have difficulty in identifying correlations when applied to sparsely sampled light curves \citep{Peterson93}, and the 7-day observed cadence does not optimally sample the time delays. Except for the V band, the DCF peaks for the UV and optical bands occur at the $\tau=0$ bin, indicating that the lags are not well resolved by the discrete sampling. In order to obtain more accurate determinations of the lags and uncertainties, we re-measured the lags using three additional methods, each of which have better sensitivity than the DCF method for detection of the  short lags in the \pks\ light curves.

\subsection{Interpolation Cross-correlation Function}
One of the most common methods for RM measurements is the Interpolation Cross-Correlation Function (ICCF), which employs a linear interpolation between successive observations \citep{Peterson04}. The ICCF quantifies the amount of similarity between two time series as a function of time shift or lag between them. We measure ICCF lags using the {\sc pyccf} code\footnote{\url{http://ascl.net/code/v/1868}} \citep{Sun18}. Since the \pks\ observations consists of $\sim$300 days with an average cadence of 3 days, we calculated the ICCF with a lag time range of $\pm$250 days and using a 0.2 day grid spacing. The resulting CCF has peak amplitude $r_\mathrm{max}$ at lag $\tau_\mathrm{peak}$. The centroid lag $\tau_\mathrm{cen}$ is determined as the centroid of all points in the CCF above $0.8r_\mathrm{max}$. 

Uncertainties were determined using the Monte Carlo flux randomisation/random subset sampling \citep[FR/RSS;][]{Peterson98, Peterson04} method, with 10,000 realisations. From the $\tau_\mathrm{peak}$ and $\tau_\mathrm{cen}$ values of each Monte Carlo realisation of the data, we obtain the cross-correlation peak distribution (CCPD) and the cross-correlation centroid distribution (CCCD), and the final values of $\tau_\mathrm{peak}$ and $\tau_\mathrm{cen}$ and their uncertainties are taken to be the median and 68\% confidence intervals of these distributions.

The ICCF lags are listed in Table~\ref{tab:lagTime}, and the CCCD is displayed in the third panel of Fig.~\ref{fig:LagJavelin}, for each combination pair between \w\ and the X-ray/UV/optical light curves. The CCCDs show narrow peaks at regularly spaced intervals corresponding to $0.5$ times the sampling interval due to aliasing, but the overall widths of these distributions are much broader than these narrow aliasing peaks, resulting in large uncertainty ranges on $\tau_\mathrm{peak}$ and $\tau_\mathrm{cen}$.  We also find a trend of decreasing $r_{\rm max}$ as a function of wavelength, similar to that observed in other \emph{Swift} RM studies \citep[e.g.,][]{Edelson19}. This can be attributed to both the lower S/N in longer wavelength bands and the increasing dilution of the AGN variability by host galaxy starlight. 


\subsection{JAVELIN}\label{sec:javelin}
\begin{figure*}
 	\includegraphics[trim=0.2cm 3.3cm 2.0cm 3.7cm, clip, width=17cm]{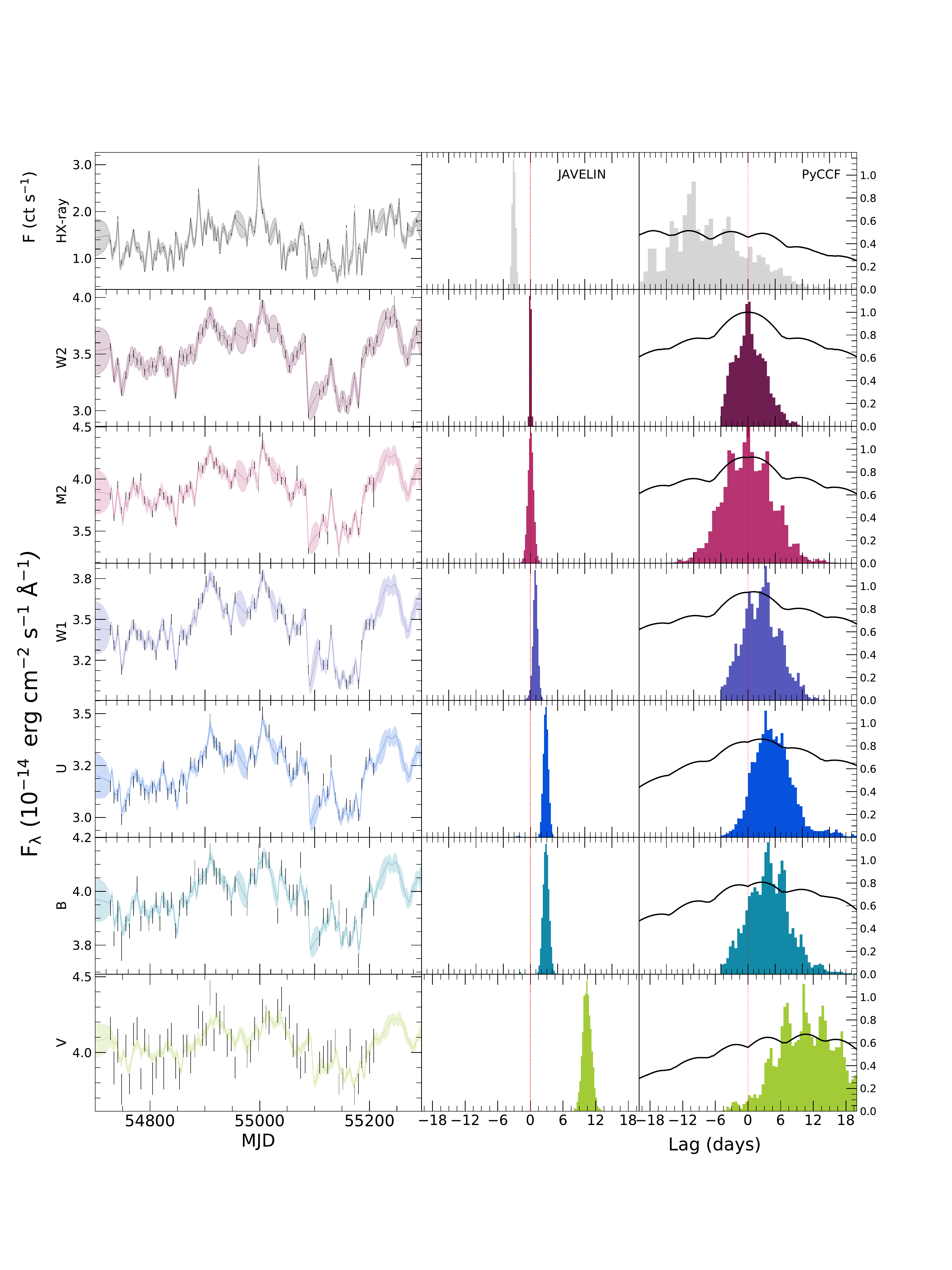}
\caption{Inter-band lag measurements of \pks. {\em Left panels:} Best fit DRW model made with {\sc javelin} to each light curve in relation to the \w-band. The contours show the 68\% confidence interval.  {\em Middle panels:} Marginalised posterior distributions of the lag measurements with {\sc javelin}. The median value and its corresponding 68\% confidence interval are shown as \tm. $\tau=0$ is shown for reference as the red vertical line. {\em Right panels:}. CCCD histograms as measured by {\sc pyccf}, the  black lines shows 0.8$r_\mathrm{max}$ used to calculate $\tau_\mathrm{cent}$. The vertical red line corresponds to $\tau=0$, for reference.
}
    \label{fig:LagJavelin}
\end{figure*}

We also used {\sc javelin}\footnote{{\sc javelin}, Just Another Vehicle for Estimating Lags In Nuclei Code, \url{https://bitbucket.org/nye17/javelin}} \citep{Zu11, Zu13, Zu16} which models the behaviour of the continuum light curve variability as an auto-regressive process using a damped-random walk model (DRW) to interpolate the reference light curve. It implements a Markov Chain Monte Carlo (MCMC) procedure via {\sc emcee} \citep{Foreman13} to sample the posterior distributions of the optimal time-delay (assuming a top-hat response function with a central value at $\tau_{\textrm{jav}}$) and scaling parameters needed for the driving light curve to match each continuum light curve. We ran the {\sc javelin} on the full light curves, with the MCMC parameters of 200 starting walkers, a 1000 step chain and burn in time of 500. 
We show the light curve together with the best fit and its $1\sigma$ confidence envelope in the left panel of Fig.~\ref{fig:LagJavelin} for each band. The marginalised posterior distributions of $\tau_{\textrm{jav}}$ are show in the middle panel of Fig.~\ref{fig:LagJavelin}, and a summary of the their median values and 68\% confidence interval are shown in Table~\ref{tab:lagTime}. 

\begin{table*}
	\centering
	\caption{Lag measurements for \pks\ obtained with 5 different methods (see Sec.~\ref{sec:measurements}) for every filter with respect to \w-band. Columns 2 and 3 give the observed and rest-frame wavelength of each filter. Column 4 lists the ICCF maximum correlation coefficient $r_\mathrm{max}$, and the uncertainty is obtained as the standard deviation of the $r_\mathrm{max}$ values from the 10,000 FR/RSS iterations. The columns 5 and 6 gives  the ICCF centroids and peaks, respectively. Columns 7 gives the lags measured by {\sc javelin}, column 8 correspond the lag obtained with {\sc mica2}. and column 9 gives the lag estimates from the  \py.}
	\label{tab:lagTime}
	\begin{tabular}{lcccccccc} 
		\hline\hline
		Band & $\lambda_\mathrm{observed}$ & $\lambda_\mathrm{rest}$ & ICCF & CCPD & CCCD & {\sc javelin} & {\sc mica2} & \py \\
		     &   &   &  $r_\mathrm{max}$ & $\tau_{\rm peak}$& $\tau_{\rm cent}$ &$\tau_{\rm JAV}$& $\tau_{\rm MICA2}$ &$\tau_{}$\\
		     &[\AA] & [\AA] &  & [days] & [days] & [days]    & [days]    &[days]   \\
		(1)     & (2) & (3) & (4) & (5) & (6) & (7) & (8) & (9) \\
		\hline
        \hline
		HX   & 3 & 2.63& 0.52$\pm 0.06$ & $-6.75^{+11.06}_{-8.68}$ & $-7.56^{+6.10}_{-8.20}$ &$-3.07^{+0.3}_{-0.4}$   & $-14.0^{+6.54}_{-6.55}$ & $\cdots$ \\[.15cm]
		UVW2 & 1928 & 1695 &1.00$\pm0.03$  &$0.00^{+0.80}_{-0.79}$ & $0.00^{+0.80}_{-0.79}$ & $0.00^{+0.11}_{-0.12}$ & $0.00^{+0.06}_{-0.05}$ & $0.00\pm0.49$ \\[.15cm]
		UVM2 & 2246 & 1975 & 0.93$\pm0.04$  & $0.80^{+4.23}_{-4.43}$ & $-0.01^{+4.23}_{-4.43}$ & $0.05^{+0.53}_{-0.53}$ & $0.05^{+0.49}_{-0.49}$ & $0.31\pm 0.61$\\[.15cm]
		UVW1 & 2600 & 2286 & 0.95$\pm0.03$ & $1.25^{+3.50}_{-3.56}$ & $2.40^{+3.50}_{-3.56}$ & $0.91^{+0.42}_{-0.43}$ & $0.82^{+0.47}_{-0.51}$ & $0.66\pm 0.74$\\[.15cm]
		U & 3467 & 3047 & 0.86$\pm0.04$  & $2.60^{+3.19}_{-3.61}$ & $4.34^{+3.19}_{-3.61}$ & $2.82^{+0.47}_{-0.46}$ & $2.21^{+0.94}_{-0.99}$ & $1.64\pm 1.10$ \\[.15cm]
		B & 4392 & 3862 & 0.80$\pm0.05$  & $2.60^{+3.90}_{-3.65}$ & $3.76^{+3.90}_{-3.65}$ & $2.86^{+0.55}_{-0.54}$ & $2.18^{+1.14}_{-1.34}$ & $2.75\pm 1.48$
		\\[.15cm]
		V & 5468 & 4808 & 0.68$\pm0.07$ & $5.98^{+5.48}_{-6.51}$ & $11.21^{+5.48}_{-6.51}$ & $10.38^{+0.82}_{-0.74}$ & $7.73^{+3.32}_{-3.28}$ & $4.06\pm1.81$ \\[.15cm]
		\hline
	\end{tabular}
\end{table*}

\subsection{MICA2}
\label{subsec:mica}
Finally, we use {\sc mica2} \citep{Li2016}\footnote{{\sc mica2} is available at \url{https://github.com/LiyrAstroph/MICA2}}, which is similar to {\sc javelin} where the variability of the driving light curve, used to interpolate, is modelled as a DRW process. A main difference is that {\sc mica2} employs a family of relatively displaced Gaussian function to model the response function instead of a top-hat. Thus, the time lag ($\tau_{\textrm{MICA2}}$) is set to be the variance-weighted centre of the Gaussian. For the sake of simplicity, we only use one Gaussian. We also test for multiple Gaussians and find the case of one Gaussian is preferable in terms of Bayesian factors. An MCMC procedure with the diffusive nested sampling algorithm (\citealt{Brewer11}) is adopted to optimise the posterior probability and determine the best estimate and uncertainties for the model parameters. In Fig.~\ref{fig:Lagmica2}, we show the best fit to the light curves (right panel) and its associated response functions (left panel). The obtained time lags are presented in Table~\ref{tab:lagTime}.

\subsection{Method comparison}
 
Overall, the lags derived from the four methods display an increasing trend with wavelength, with values consistent across methods for the UV optical bands albeit larger uncertainties.  These values suggest that the variations shown by \pks, follow the expected trend of larger lags at longer wavelengths in stark contrast to the results reported by \citet[][see Section~\ref{sec:discussion} for further details]{Gli13}. 

\begin{figure}
     \includegraphics[trim=1.cm 1.2cm 1.2cm 2cm,clip, width=\columnwidth,angle=0]{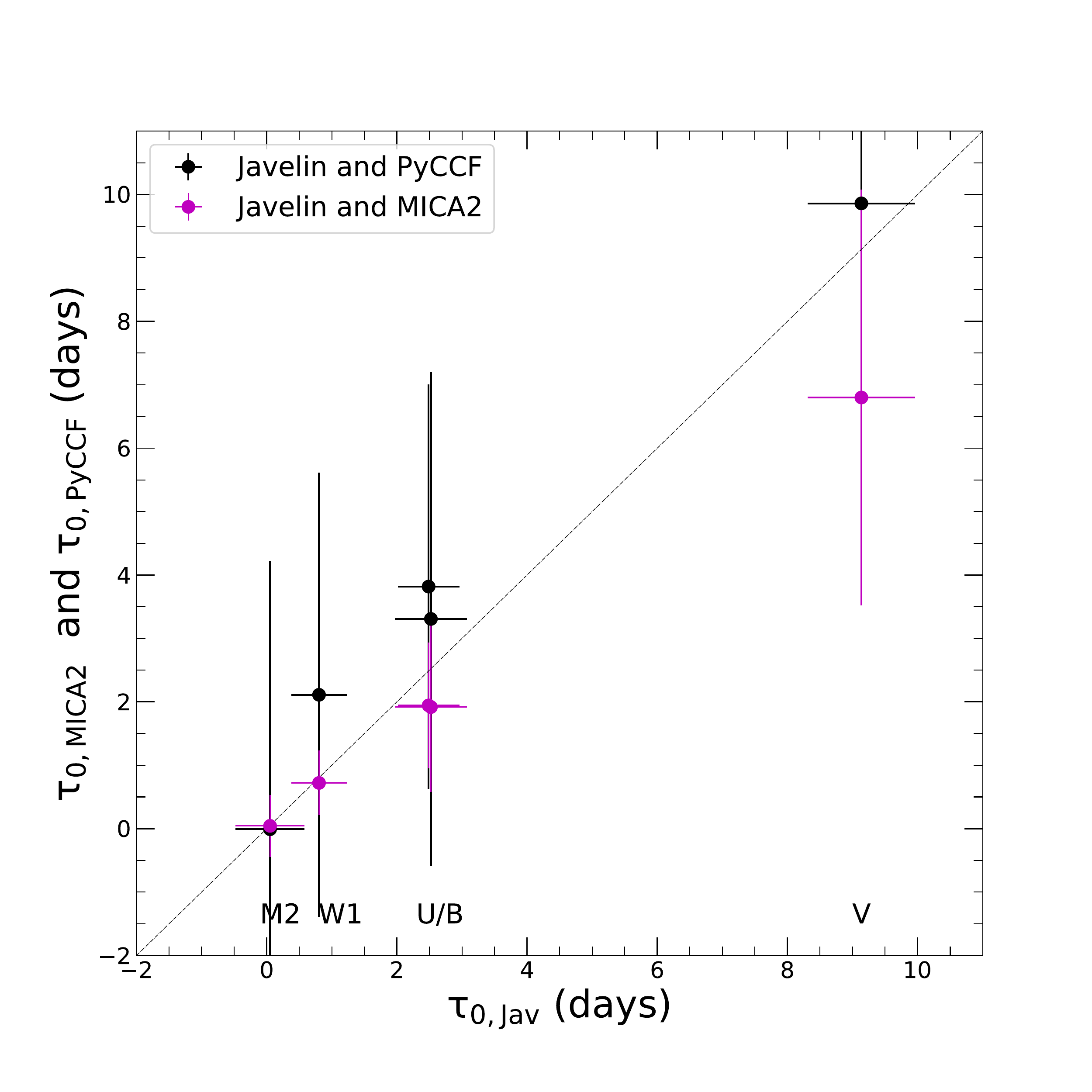}
\caption{Lag measurement comparison between methods: {\sc javelin} vs {\sc pyccf} (black), and {\sc javelin} vs {\sc mica2} (purple). The solid line shows the one-to-one relation for reference.
}
    \label{fig:comparison}
\end{figure}

In Fig.~\ref{fig:comparison}, we show a comparison between the measurements obtained with three additional time analysis methods {\sc javelin} against {\sc mica2} (purple) and {\sc pyccf} (black). These lag measurements are consistent with each other within 1$\sigma$ for all bands (all results are presented in Table~\ref{tab:lagTime}). The median lag values obtained with {\sc pyccf} (using the ICCF method) are similar to those obtained with {\sc javelin} and {\sc mica2} albeit larger uncertainties (see Fig.~\ref{fig:comparison}) due to the ICCF method being less sensitive in detecting lags with low cadence light curves and lower S/N \citep{Li19}. Therefore, we have adopted the results obtained with {\sc javelin} and {\sc mica2} as the values used in the analysis in the following sections.

\begin{figure}
     \includegraphics[trim=0cm 0cm 0cm 0cm,clip, width=\columnwidth,angle=0]{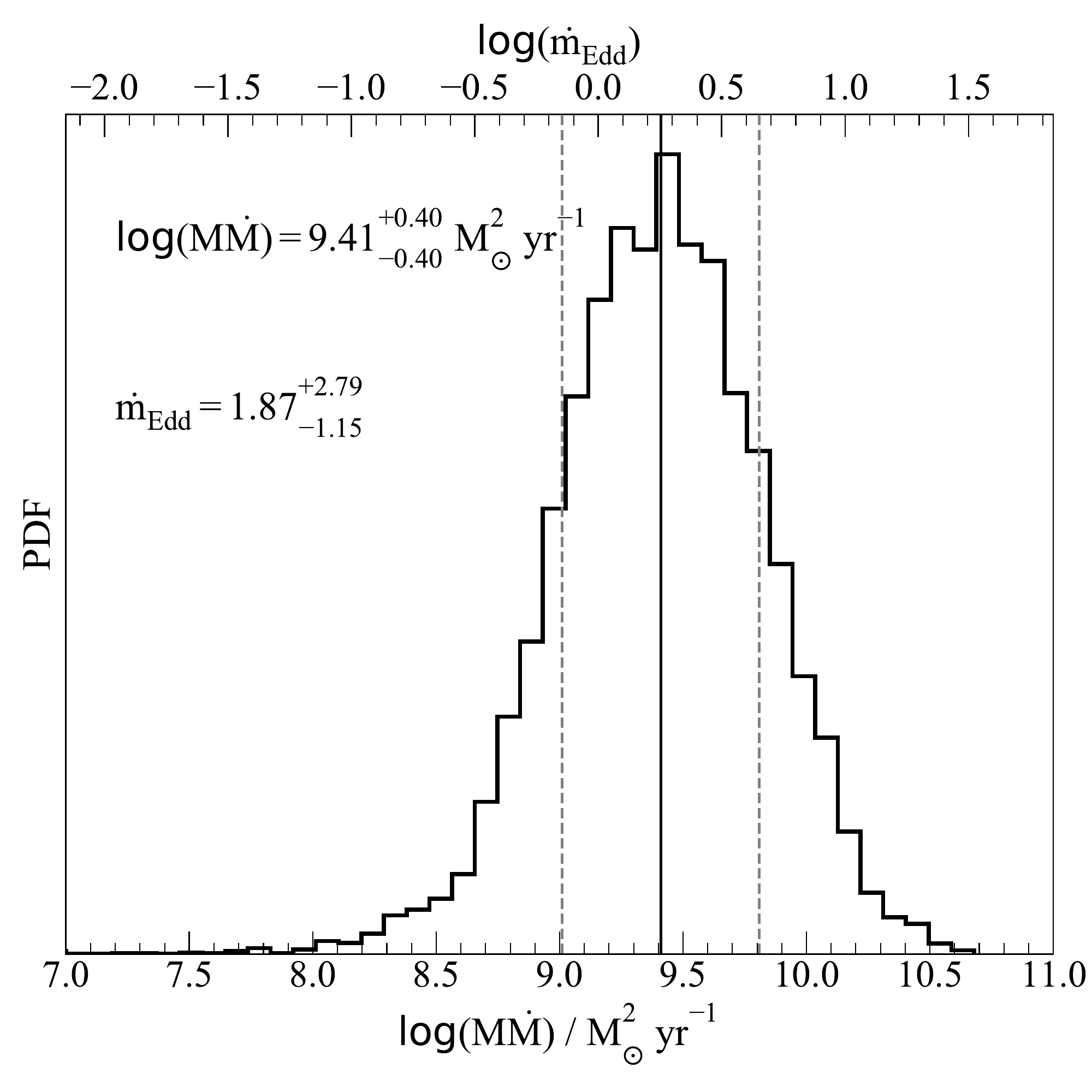}
\caption{The marginalised posterior distribution for the product $M\dot{M}$ as inferred by \py\ (see Sec.~\ref{sec:cream}), with the median and 1$\sigma$ marked as solid and dashed lines, respectively. We calculate the Eddington ratio shown in the top axis, assuming a BH mass of $2.5\times10^8$ M$_{\odot}$, and a fixed inclination angle of $i=0^\circ$ and a temperature profile $T\propto R^{-3/4}$ within the PyceCREAM fit.
}
    \label{fig:mdotcream}
\end{figure}

\section{Accretion disc modelling}\label{sec:cream}
We also analyse the light curves using the Continuum REprocessed AGN Chain Monte Carlo \py\ code described in \citep{Starkey16}\footnote{We used the {\sc python} wrapper \py\ \url{https://github.com/dstarkey23/pycecream}} to infer the properties of the accretion flow in \pks. Here, we present a brief description of the model and refer to \citet{Starkey16,Starkey:2017} for further details of the algorithm. \py\ assumes that 
the variability of the continuum light curves are described within the reprocessing model. A variable X-ray source (corona) located above the supermassive black hole at a height of the accretion disc acts as a ``lamp-post'' that shines and heats up the disc \citep{Cackett2007}. This additional energy input on the viscously heated disc is in turn thermalised and re-emitted at longer wavelengths. This X-ray reprocessing produces correlated variations in the emission of the disc which propagates radially outwards, and produces ``light echoes''  exciting first the variations of the hot internal disc and then the variations of the external part of the cold disc. 
The wavelength-dependent variations probe different regions of the disc and allow to measure the temperature profile $T(R)$, where $T(R)\propto R^{-3/4}$ corresponds to a standard accretion disc. Therefore, the expected time delay at a given the observed wavelength (which probes a characteristic temperature/radius on the disc) scales as $\tau=R/c\propto(M_{BH}\dot{M})^{1/3}T^{-4/3}\propto(M_{BH}\dot{M})^{1/3}\lambda^{4/3} $, where  $M_{BH}$ is the mass of the black hole and  $\dot{M}$ is the rate of mass accretion. Thus, the delay distribution at different wavelengths inferred via the light-curves carries information of the product $M_{BH}\dot{M}$. 

Here, we have assumed the quasar to be face-on, i.e. $i=0$, and set the temperature profile index to $-3/4$. The effect of fixing the inclination has a negligible effect on the measured average delay, as shown by \citet{Starkey16}. As for the driving light curve, PyceCREAM models it as a Fourier time series using a random-walk prior to constrain the power density spectrum. We used a fixed high frequency cutoff for the power spectrum of 0.5 cycles d$^{-1}$. Then, each light curve is shifted and stretched to match the flux levels at each wavelength (after the convolution with the delay distribution). We used uniform priors for all parameters except for the driving light curve power spectrum. We ran the MCMC sampling for $10^{5}$ iterations, discarding the first third of the chains as the burn-in phase. 

In Figure~\ref{fig:creamJuan}, we show the best fit to the UV and optical light curves, as well as the delay distribution for each band. The driving light curve, shown in the top most panel, provides a good description of the variability at all wavelengths.
The mean delays for each band are consistent with those obtained from other methods (see Table~\ref{tab:lagTime}) following --by construction-- the $\tau\propto\lambda^{4/3}$ relation. The uncertainties in the mean lags are larger than those obtained by other methods. This likely arises from the flexibility of MICA and Javelin to fit every band independently. On the other hand, PyceCREAM fits all bands simultaneously and restricts the lag measurements to follow the $\tau\propto\lambda^{4/3}$ relationship. As a consequence, any scatter around this relationship line will introduce additional variance scatter in the physical parameters inferred by PyceCREAM and thus, in the mean lags.
Furthermore, we find a value for the product $\log( M_{BH}\dot{M}$ / M$_{\odot}^2$ yr$^{-1}$) = $9.4\pm0.4$; the marginalised posterior distribution is shown in Fig.~\ref{fig:mdotcream}. Using the BH mass of $2.5\times10^8$ M$_{\odot}$, we estimate a mass accretion rate (divided by its Eddington limit) of $\dot{m}_{\textrm{Edd}}=1.87^{+2.79}_{-1.15}$. This value is above its Eddington limit consistent with previous measurements $\dot{m}_{\textrm{Edd}}=1.7$ \citep{Gli10}.

\section{Discussion}
\label{sec:discussion}
\citet{Gli13} presented the results of the first continuum RM study of \pks\, with {\em Swift}. 
The inter-band lag measurements, made using the DCF method, found a trend of negative lag times towards the UV and optical bands, while positive values were found towards the X-ray bands. This result was interpreted as accretion induced fluctuations moving inwards through the disc \citep{Lyubarskii97, Arevalo08}  where the optical bands are the driver of the observed changes in the UV bands, and in turn, responsible for those observed in X-rays. This trend is the opposite of the disc reprocessing model or ``lamp-post". This model postulates that the corona and/or internal part of the disc illuminates and heats the outer regions of the disc by disturbing its local temperature, thus generating wavelength-dependent delays increasing at longer wavelengths \citep{SS73, Frank02, Cackett2007}.

In this work, we revisited the lag measurements of the \textit{Swift}/UVOT light curves taken from \citet{Gli13}. We recreated the DCF analysis presented in \citet{Gli13} and found that their measurements must have inadvertently used the reference W2 band as the responding light curve rather than the driving one. This was reflected as a mirror image cross-correlation function and thus provided an opposite sign in the measurement of the inter-band delays (see Fig.~\ref{fig:dcf}). Further analysis with {\sc javelin}, {\sc mica2} and ICCF/{\sc pyccf} codes (described in Section~\ref{sec:measurements}) confirm the wavelength-dependent trend in which the longer wavelength light curves lag the shorter wavelength ones (see Fig.~\ref{fig:LagJavelin} and Table~\ref{tab:lagTime}), in line with the disc reprocessing model and most continuum reverberation studies. We show that the UV and optical light curves are well correlated and the lag time measurements obtained with four different codes/methods are widely consistent with each other. 

\subsection{Lag spectrum analysis }
\label{sec:fit}
\begin{figure*}
     \includegraphics[trim=0cm 1.6cm 0.5cm 1cm 1cm, clip,width=18cm,height=3in]{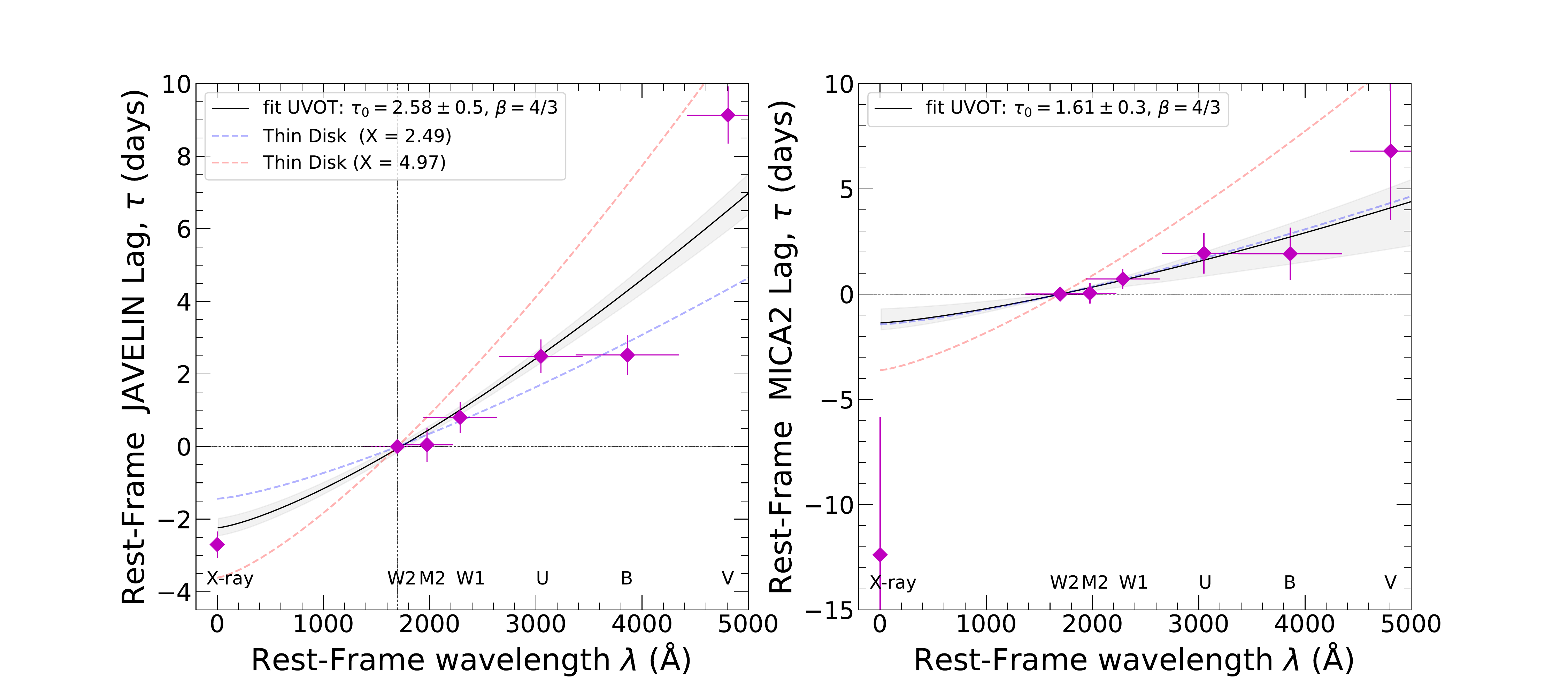}
 \caption{Delay spectrum of \pks\ using the \w\ band as reference. The {\sc javelin} (left-panel) and {\sc MICA2} (right-panel) measurements (purple circles, transformed to the AGN rest frame) show an increasing trend with wavelength. The best fit (black and red line, with shaded envelope illustrating the uncertainty range) follows the reprocessing model relation of $\tau\propto\lambda^{\beta}$, with a fixed power-law index $\beta=4/3$. For {\sc JAVELIN} data, the X-ray data was not included in the fit (black line) of the figure. For the fit (red line) shown in the {\sc MICA2} figure all data were included.  The red and blue dotted line correspond to the predictions for a thin-disk model using the equation~\ref{eq:3} considering the value of $X=2.49$ (blue line) and $X=4.97$ (red line).
     }
     \label{fig:Lagfit}
\end{figure*}

We can use the delay spectrum for \pks, as measured with {\sc javelin} (see Table~\ref{tab:lagTime}) to test the predictions from accretion disc theory. We transformed the delay measurements and wavelengths to the AGN rest frame. We then proceeded to fit the delay spectrum with the functional form:
\begin{equation}
\label{eqn:cont_lag}
\tau = \tau_{0}\left(\frac{\lambda}{\lambda_{0}}\right)^{\beta} - y_{0}\,,
\end{equation}
where $\lambda_{0}$ is the reference wavelength corresponding to the rest-frame \w-band ($\lambda 1695.4$ \AA), $\tau_{0}$ is the reference time which measures the radius of the disc emitting at the reference wavelength $\lambda_{0}$,  $\beta$ is the power-law index which reflects the disc temperature profile, and $y_0$ allows the model to pass through zero lag at the reference wavelength $\lambda_{0}$. We made a fit to the lag time data obtained with {\sc JAVELIN} and {\sc MICA2} (see~\ref{fig:Lagfit}) using the Eq.~\ref{eqn:cont_lag} and considering the $\beta$ parameter fixed at 4/3 as predicted by the standard thin-disc theory \citep{SS73} and leaving $\tau_0$ free. The fit was applied twice: first to the set of all lag-time measurements from X-rays to the V-band, and a second fit was then applied to the UVOT data only. For {\sc JAVELIN}, we find that the fits for the two models are very similar with values of $\tau_{0}=2.69\pm0.40$ days with $\chi^2_\nu=4.75$ considering all points, while excluding the X-ray data gives a value of $\tau_{0}=2.58\pm0.50$ days and a $\chi^2_\nu$ of 5.63. For the {\sc MICA2} case, the values obtained for $\tau_{0}$ are $1.66\pm044$ days for X-ray+UVOT data and $1.61\pm0.29$ days for UVOT data with $\chi^2_\nu$ of 0.86 and 0.37 respectively. The summary of the results are listed in table~\ref{tab:lagfitinfo} and Figure~\ref{fig:Lagfit} shows the lag times for each band (purple circles) together with the fit made to the UVOT data (black line) and Xray+UVOT data (red line). 

The $\beta=4/3$ model follows the general trend of the data, although the B and V-band points scatter significantly below and above the best-fit model respectively. The corresponding high $\chi^2_\nu$ values obtained with {\sc JAVELIN} indicate that the power-law model for the lag spectrum is not formally a good fit for these data. The $\chi^2_\nu$ value obtained with {\sc MICA2} is more reasonable, indicating that the reprocessing model is an acceptable match to the data. However, we note that the {\sc JAVELIN} measurements for the B and V bands carry substantially smaller uncertainties than the lags measured using the other methods ({\sc PyCCF}, {\sc MICA2}, and {\sc PyceCREAM}), and the interpretation of any model fit is subject to considerable ambiguity given the different results obtained from different lag measurement techniques.

\begin{table*}
	\centering
	\caption{Parameters for lag-wavelength fits. Column 2: shows the filters used for the fit, all-data corresponds to the X-ray data up to the V-band, and UVOT corresponds to the \w\, to V bands.  columns 3,4,5 correspond to the parameters obtained from the fit of equation~\ref{eqn:cont_lag}. Column 6: $\chi^2$/degrees of freedom.}
	\label{tab:lagfitinfo}
	\begin{tabular}{llcccc} 
		\hline\hline
		 Code  &Data & $\tau_0$ & $\beta$ & $y_{0}$ & $\chi_\nu^2$\\
               &   & (days)          & &    \\  
         (1)  &  (2)   & (3) & (4) & (5) & (6) \\      
         \hline
         {\sc JAVELIN}& & &  &  & \\
		 &X-ray + UVOT & 2.69$\pm$0.40 & 4/3 & 0.88$\pm$0.08 & 4.75 \\
		 &UVOT & 2.58$\pm$0.50 &  4/3 & 0.86$\pm$0.09 & 5.63\\
		 \hline
		 {\sc MICA2}& & &  &  & \\
		  & X-ray + UVOT & 1.66$\pm$0.44 & 4/3 & 0.84$\pm$0.03 & 0.86\\
		  & UVOT & 1.61$\pm$0.29 & 4/3 & 0.84$\pm$0.02 & 0.37 \\
\hline\hline
	\end{tabular}
\end{table*}

We also  compare the observed lag-wavelength behaviour against theoretical expectations to probe the size of the accretion disc. We followed the methodology proposed by \citet{Fausnaugh16} and \citet{Edelson17}, to estimate the expected photon travel time $r(\lambda)/c$ (where $r$ is the disc radius at wavelength $\lambda$) from the inner part of the accretion disc to the outer region, using the following equation:

\begin{equation}
    r(\lambda) = 0.09\left(X\frac{\lambda}{\lambda_{0}} \right)^{4/3} M_{8}^{2/3} \left( \frac{\dot{m}_\mathrm{Edd}}{0.10}\right)^{1/3} \mathrm{lt-days}\,,
    \label{eq:3}
\end{equation}
where $\lambda_0$ corresponds to the rest-frame wavelength of the driving light curve (\w), $M_{8}$ is the black hole mass in units of $10^8\,M_{\sun}$, and $\dot{m}_{\rm Edd}$ is the Eddington ratio $L_{\rm bol}/L_{\rm Edd}$.  
The multiplicative factor $X$ incorporates the geometry of the delay distribution. Following \cite{Edelson19} we consider two cases. If we simply assume that at radius $r$ the disc emits at a wavelength corresponding to the local temperature via Wien's law, then the scaling factor is $X=4.97$. For a model in which $r$ corresponds to the flux-weighted radius for a disk emitting locally as a blackbody, then $X=2.49$.
To obtain estimates of $r(\lambda)$ across the wavelength range of the data, we have considered both $X$ values, a mass of the BH of $2.5\times10^{8}~M_{\sun}$, and the accretion rate $\dot{m}_{\rm Edd}=1.7$ \citep{Gli10}.  

In Figure \ref{fig:Lagfit} we plot curves of $[r(\lambda) - r(\mathrm{W2})]/c$ for these two values of $X$, to compare with the observed lag-wavelength curve. We find that for the {\sc JAVELIN} data the best-fitting 4/3 model is found between the two model curves (black line - left panel), while the best fit found for the {\sc MICA2} data (red line - right panel) falls exactly on the  flux-weighted  model (X=2.49). 
This might be naively interpreted as an indication that the disc in \pks\ is $\sim50\%$ larger than expected according to the {\sc JAVELIN} results or has a size of $1.66\pm0.4$ lt-days if we consider the {\sc MICA2} results and take its black hole mass and Eddington ratio. However, the data points from the UV through B bands are largely compatible with the ($X=2.49$) model within their uncertainties, and the discrepancy between this model and the data is almost entirely the result of the long lag measured for the V band. In some other objects having high-quality UV-optical monitoring data, the disc radius exceeds model predictions by a factor of $\sim3$ \citep[e.g.,][]{Fausnaugh16}. While it is possible that the disc radius of \pks\ is also in excess of model predictions, albeit by a smaller factor, the data quality in this case is not sufficient to draw unambiguous conclusions. Furthermore, any comparison between the disc model predictions and data will be subject to the usual uncertainties in the estimated black hole mass and Eddington ratio, which may be of order $\sim0.5$ dex and certainly exceed the observed discrepancies between the model predictions and the data.

Another question of interest is whether the data show an excess lag in the U band that would indicate a substantial contribution of Balmer continuum emission from the broad-line region \citep{Korista01,Chelouche:2019,Netzer:2020}. While definite excess U-band lags have been observed in \emph{Swift} data of some other sources \citep[e.g.,][]{Fausnaugh16, Edelson19}, the data in this case do not give a clear result. The U-band lag as measured by the ICCF method exceeds the the B-band lag by $\sim0.6$ days, but the difference is much smaller than the measurement uncertainties, and {\sc JAVELIN} and {\sc MICA2} each obtain nearly equal lags in the U and B bands with substantial uncertainties. In this case, the weekly observing cadence is not optimal for discerning the details of wavelength-dependent lag behaviour on timescales less than several days.

Despite these limitations, the reverberation measurements nevertheless lead to the clear conclusion that PKS 0558--504 does follow an increasing trend of lag against wavelength similar to that observed in other intensively monitored AGN. Further monitoring of this object, at a more rapid cadence and ideally extending to wavelengths longer than the V band, would provide a basis for more detailed and rigorous comparison with model predictions.

\section{Conclusions}
\label{sec:conclusion}
In this paper, we present the results of a revision to measurements of continuum reverberation lags for the AGN \pks\, based on \emph{Swift} observations that were carried out during 2008--2010. This object had previously been an outlier among AGN having \emph{Swift} UV-optical continuum reverberation data, in that \citet{Gli13} found that the UV variations lagged behind the optical variations, and the X-rays lagged behind the UV. This trend is opposite to the behaviour found in other AGN with intensive monitoring data from \emph{Swift}: in other Seyferts, the lags are observed to increase as a function of wavelength. Based on their result, \citet{Gli13} argued against the disc reprocessing model for the variability. The puzzling behaviour found for \pks\ prompted us to review the lag measurement for this object. Applying the same DCF method to reproduce the \citet{Gli13} measurement, we found that the data actually followed the more canonical trend of optical wavelengths lagging behind the UV, and UV lagging behind the X-rays.

To examine the reverberation lags more closely, we carried out new measurements with four codes: {\sc pyccf}, {\sc javelin}, {\sc mica2}, and \py, taking the \w\ band ($\lambda_\mathrm{obs} = 1928$~\AA) light curve as the driving band. We find that the variations of the UV/optical light curves are strongly correlated, but there is a poor correlation with the X-ray bands (see Fig.~\ref{fig:LagJavelin}). The results obtained with all four codes show similar delay spectra, with a clear trend of increasing lags as a function of increasing wavelength. All methods demonstrate the opposite lag-wavelength trend from that reported by \citet{Gli13}. Our results demonstrate that \citet{Gli13} must have inadvertently swapped the ordering of the driving and responding light curves when calculating the DCF.

We present fits to the delay spectrum (Fig.~\ref{fig:Lagfit}) using the relation $\tau\propto\lambda^{\beta}$ for $\beta$ fixed at 4/3 (Eq.~\ref{eqn:cont_lag}), expected for an optically-thick and geometrically-thin accretion disc \citep{Cackett2007}. While the $\beta=4/3$ model appears to be compatible with the lag-wavelength trend in the data, the large uncertainties in the data precludes  us from exploring deviations from the canonical temperature-radius (which sets $\beta$). We also compared the data with the standard model prediction for the disc radius as a function of emitting wavelength to test whether the disc size is too big, as seen in other continuum reverberation mapping experiments \citep[e.g.,][]{Edelson19}. Using a model for the flux-weighted radius as a function of wavelength, we find that the observed lags indicate a continuum emission region $\sim50\%$ larger than predicted by the disc model. If we consider the {\sc JAVELIN} results, or the \pks\ disk has a size similar to that predicted by  model if we consider the {\sc MICA2} result.  For the two results, the values of its black hole mass and Eddington ratio were taken.
However, given the anticipated uncertainties in these parameters as well as the scatter in the measured lags, the apparent discrepancy between the disc model and the data is not significant. Nevertheless, the data indicate that \pks\ is a promising subject for additional monitoring observations, and further UV through optical monitoring at a daily cadence would be able to resolve the optical lags well and provide a definitive measurement of the wavelength-dependent lag behaviour in this object.

\section*{Acknowledgements}
DHGB acknowledges CONACYT support \#319800 and of the researchers program for Mexico. Research by AJB is supported by National Science Foundation grant AST-1907290. JVHS acknowledges funds from a Science and Technology Facilities Council grant ST/R000824/1 research fellowship.  YRL acknowledges the financial support from the National Natural Science Foundation of China through grant No. 11922304 and from the Youth Innovation Promotion Association CAS.

\section*{Data availability}

All of the data used in this work are available in the Supplementary Data section of \citet{Gli13} at \url{https://academic.oup.com/mnras/article/433/2/1709/1751179}.

\bibliography{bibliography} 
\bibliographystyle{aasjournal}


\appendix

\section{Lag measurement with MICA2}

\begin{figure*}
 	\includegraphics[trim=0cm 0.1cm 0cm 0.1cm 1cm, clip,width=18cm]{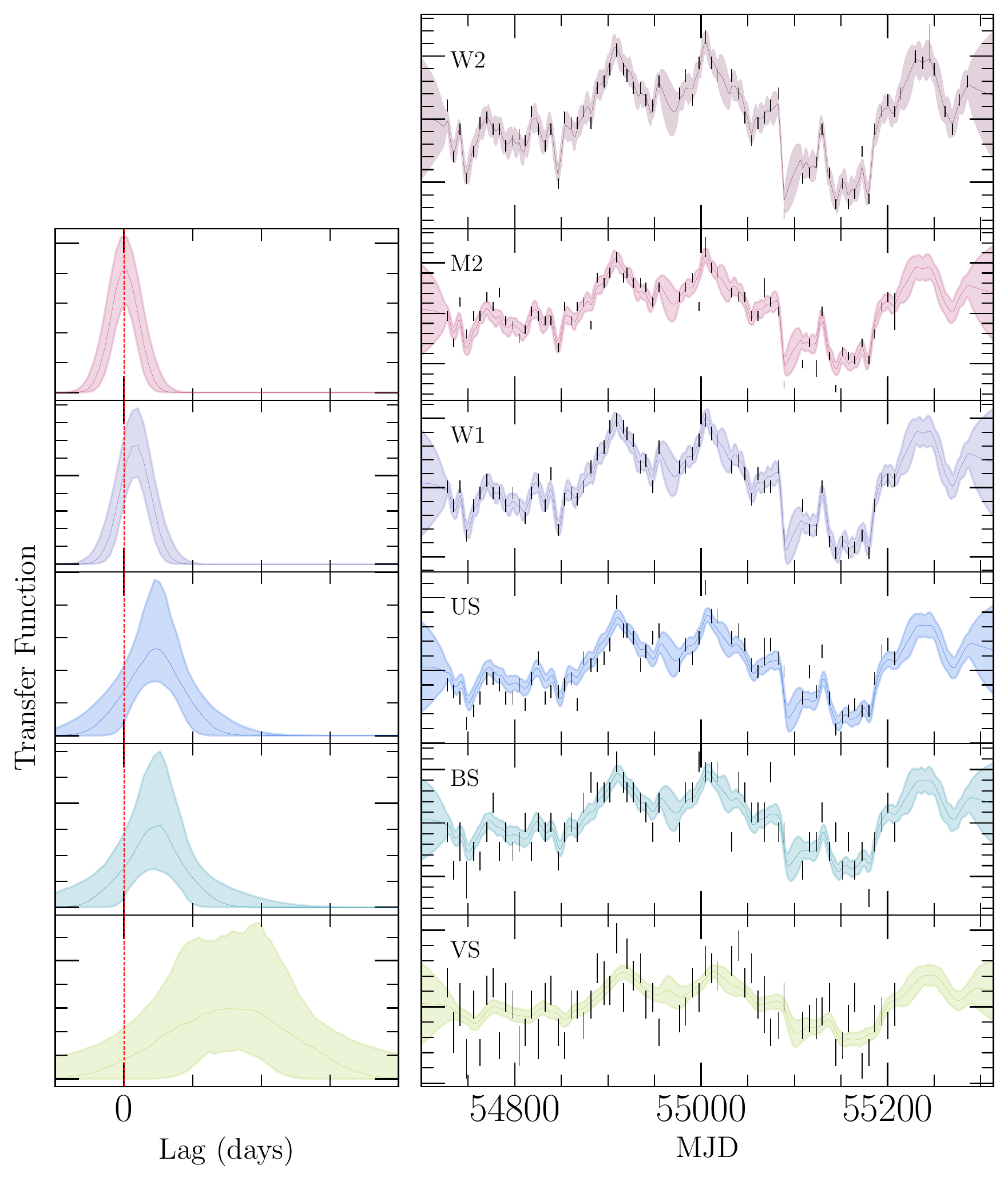}
\caption{Inter-band lag measurement performed with {\sc mica2}. The top panel shows the \w reference light curve used to fit the rest of the UVOT filters. Left panels show the transfer function for each band, in respect to the UVW2 band. Right panel shows the photometry (blue) with the best fit model (black line), with the grey envelopes displaying the 1$\sigma$ confidence interval. }
    \label{fig:Lagmica2}
\end{figure*}

Figure~\ref{fig:Lagmica2} shows the \pks\, light curves obtained with the six UVOT filters, together with the lag time measurements and fits made with the {\sc mica2} code. For more details see section~\ref{subsec:mica}.

\section{Reverberation modelling with \py}

\begin{figure*}
 	\includegraphics[trim=0cm 3.5cm 2cm 1cm, clip, width=17cm]{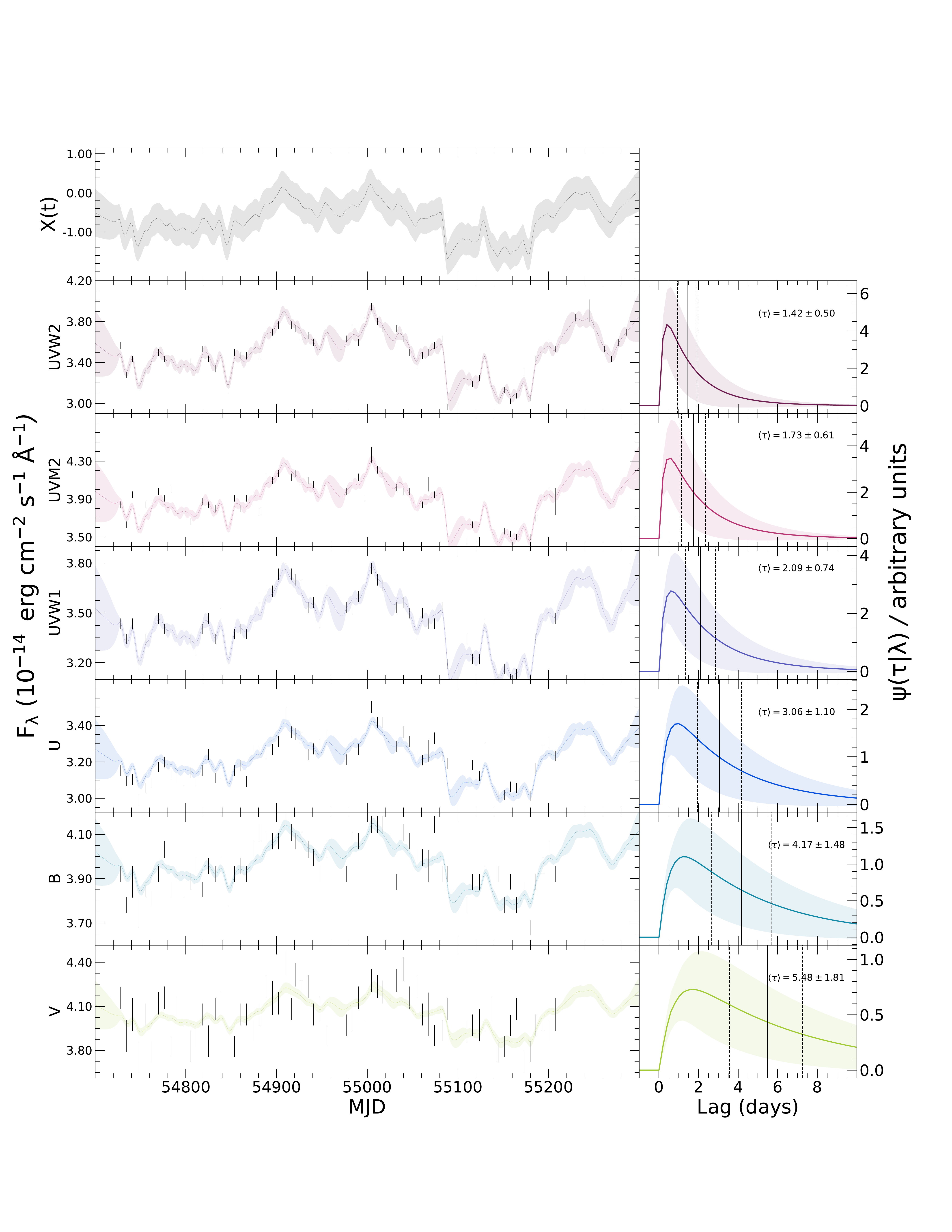}
\caption{ Reverberation model fit with \py\ to a face-on accretion disc and a $T\propto R^{-3/4}$ temperature profile for \pks. The top panel shows the inferred driving light curve. Right panels show the delay distribution for each band. The black vertical lines correspond to the mean lag time $\langle \tau \rangle$ and the dashed vertical lines correspond to 16\% and 84\% persentile.. Left panel shows the photometry with the best fit model. The black points around the dotted grey line show the residuals. All envelopes represent the 1$\sigma$ confidence interval.}
    \label{fig:creamJuan}
\end{figure*}

Figure~\ref{fig:creamJuan} shows the best fit from \py\ to each Swift band.

\label{lastpage}
\end{document}